\documentclass[pr1,amsmath,twocolumn,showkeys,superscriptaddress,preprintnumbers]{revtex4}

\usepackage{gensymb}
\usepackage{graphicx,color}
\usepackage{amssymb}
\usepackage{enumerate}
\usepackage{verbatim}
\usepackage{natbib}
\usepackage{amsmath}

\begin{document}
  \newcommand {\nc} {\newcommand}
  \nc {\Sec} [1] {Sec.~\ref{#1}}
  \nc {\IR} [1] {\textcolor{red}{#1}} 
  \nc {\IB} [1] {\textcolor{blue}{#1}}
  \nc {\CGMF}{$\mathtt{CGMF}$}

\title{Prompt Neutron Multiplicity Distributions Inferred from $\gamma$-ray and Fission Fragment Energy Measurements}

\author{A.~E.~Lovell}
\email{lovell@lanl.gov}
\affiliation{Los Alamos National Laboratory, Los Alamos, NM 87545, USA}
\author{I.~Stetcu}
\affiliation{Los Alamos National Laboratory, Los Alamos, NM 87545, USA}
\author{P.~Talou}
\affiliation{Los Alamos National Laboratory, Los Alamos, NM 87545, USA}
\author{G.~Rusev}
\affiliation{Los Alamos National Laboratory, Los Alamos, NM 87545, USA}
\author{M.~Jandel}
\affiliation{Los Alamos National Laboratory, Los Alamos, NM 87545, USA}
\affiliation{Department of Physics and Applied Physics, University of Massachusetts Lowell, Lowell, MA 01854, USA}

\date{\today}


\begin{abstract}
	We propose a novel method to extract the prompt neutron multiplicity distribution, $P(\nu)$, in fission reactions based on correlations between prompt neutrons, $\gamma$ rays, and fragment kinetic energy arising from energy conservation.  In this approach, only event-by-event measurements of the total $\gamma$-ray energy released as a function of the total kinetic energy (TKE) of the fission fragments are performed, and no neutron detection is required.  Using the {\CGMF} fission event generator, we illustrate the method and explore the accuracy of extracting the neutron multiplicity distribution when taking into account the energy resolution and calibration of the energy measurements.  We find that a TKE resolution of under 2 MeV produces reasonably accurate results, independent of typical $\gamma$-ray energy measurement resolution.
\end{abstract}

\preprint{LA-UR-19-24471}

\keywords{Fission, prompt neutrons, prompt $\gamma$ rays, correlated observables}

\maketitle

\section{Introduction}

Accurately describing the fission process is important for a variety of applications including nuclear non-proliferation, stockpile stewardship, and energy production, as well as for a fundamental understanding of nuclear physics.  For many applications, knowledge beyond average quantities, such as average neutron and $\gamma$-ray energy and number of prompt particles emitted per fission event, is necessary, e.g. \cite{Talou2018}.  Knowledge of detailed distributions is also critical, such as neutron and $\gamma$ multiplicities and energy spectra, along with the correlations between all observables.  

During the past several decades, a plethora of experiments measuring fission fragment yields, angular correlations, prompt particle energies, and multiplicity distributions have been conducted on fissile nuclei, in particular for the spontaneous fission of $^{252}$Cf and neutron-induced fission of $^{235}$U and $^{239}$Pu (e.g. \cite{Talou2018} and references therein).  Measurements for a large number of observables in all major and some minor actinides exist for spontaneous fission and thermal neutron-induced fission but are increasingly scarce as incident neutron energies increase.

Most experimental setups are designed to detect either neutrons \cite{Devlin2018} or $\gamma$ rays \cite{Ullmann2013} or fission fragments \cite{Pellereau2017,Meierbachtol2015,Fregeau2016,Heffner2014} but rarely to measure correlated data \cite{Marcath2018,Pozzi2014}.  When validating a model for correlated observables, it is often necessary to rely on separate measurements from different facilities and experimental setups.  Some correlations between observables are very important and are well known.  The emission of prompt neutrons from the fission fragments strongly depends on the excitation energy available in each fragment.  The average number of neutrons, or multiplicity, decreases with increasing kinetic energy of the two fission fragments \cite{Gook2014,BudtzJorgensen1988,Bowman1963}.  On the other hand, some of these correlations remain the subject of disagreement, such as the correlations between prompt neutron and $\gamma$-ray multiplicities \cite{Nifenecker1972,Glassel1989,Wang2016,Bleuel2010,Marcath2018}.  In addition, the direct measurement of neutrons is technically challenging, especially as the incident neutron energy increases due to neutron scattering in and around the detector setup.  For neutron-induced fission on $^{235}$U, $^{238}$U, and $^{239}$Pu, there is only one measurement of the neutron multiplicity distribution for incident neutron energies above thermal \cite{Zucker1986}.

Due to energy conservation, the number and energies of the emitted neutrons should be correlated with the excitation energy of the decaying fragment and the energy of the emitted $\gamma$ rays.  
In this paper, we propose a novel method to exploit energy correlations between the prompt $\gamma$ rays emitted in fission and the total kinetic energy (TKE) of the fission fragments in order to extract the multiplicity distribution of the prompt fission neutrons, $P(\nu)$.
 


\section{Methods}

Our theoretical study relies on the {\CGMF} code \cite{CGMF,Becker2013,Stetcu2014}.  The fission fragments are treated as compound nuclei that release their energy through successive emissions of neutrons and $\gamma$ rays from the fully accelerated fragments.  Distributions in mass, charge, total kinetic energy, spin, and parity are necessary inputs and used as initial conditions for the decay.  Both $Y(A)$ and $Y(TKE|A)$ are Gaussian, traditionally obtained from systematics constrained by available experimental data (e.g. \cite{stetcu-nd2016,stetcu-cnr15}), and the charge distributions are taken as a function of incident energy from the Wahl systematics \cite{Wahl2002}.  Once the yields are sampled, the total excitation energy available in the fragments is determined from energy conservation and shared between the two fragments through a ratio parameter, $R_T$.  This ratio is taken to be a function of fragment mass and is estimated to reproduce the average number of neutrons emitted by each fragment.  The energy sharing, in particular, is important for reproducing prompt neutron and $\gamma$-ray observables. The distribution of spins of the daughter fragments influences the competition between neutron and $\gamma$-ray emissions, described by $P(J) \propto (2J+1)\mathrm{exp}\left [ -J(J+1)/2B^2(Z,A,T) \right]$, where $B^2$ is the spin-cut off parameter \cite{Stetcu2014}.  The width of this distribution can be tuned using a global multiplying factor of the spin-cut off parameter, $\alpha$.  The emission of the neutrons and $\gamma$ rays is modeled in a Monte Carlo implementation of the Hauser-Feshbach statistical theory of nuclear reactions \cite{HauserFeshbach1952}.  The complete history of all fission fragment, neutron, and $\gamma$-ray momenta are recorded, which are then used to infer a wide range of correlations.



In this work, we study the fission of three isotopes:  $^{252}$Cf spontaneous fission and neutron-induced fission of $^{235}$U and $^{239}$Pu at thermal and 4.0 MeV incident neutron energies.  For each nucleus and energy, the calculations from {\CGMF} are used to obtain correlations between the total $\gamma$-ray energy, $E^\mathrm{tot}_\gamma$, and the total kinetic energy, TKE, on an event-by-event basis, as well as for the average total $\gamma$-ray energy, $\overline{E_\gamma^\mathrm{tot}}$, as a function of TKE.  Here, we consider the values of TKE before neutron emission, $\mathrm{TKE}_\mathrm{pre}$, and after neutron emission, $\mathrm{TKE}_\mathrm{post}$. 

Energy conservation provides a direct correlation in each fission event given by
$\mathrm{TXE} = Q - \mathrm{TKE}_\mathrm{pre},$
\noindent for the $Q$-value of the reaction, $Q$, and the total excitation energy, TXE, shared between the two fission fragments, assuming that no neutrons are emitted before scission, at scission, or during the acceleration process.  For each event, TXE is the sum of $E^\mathrm{tot}_\gamma$ and the total neutron energy, $E^\mathrm{tot}_n$.

\begin{figure*}
	\centering
	\includegraphics[width=\textwidth]{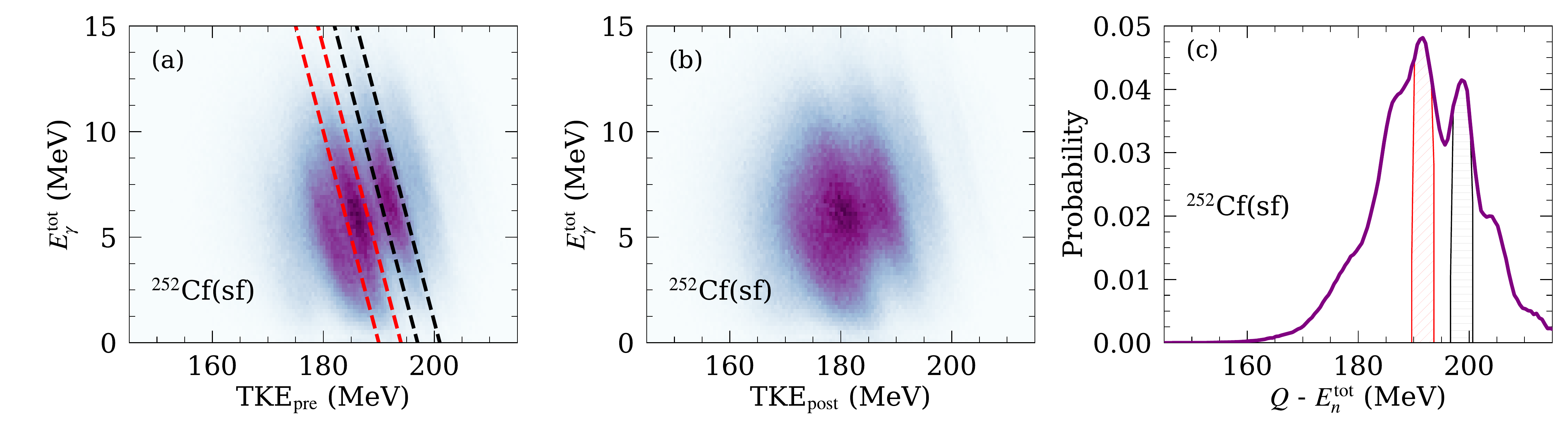} 
	\caption{(Color online) Total $\gamma$-ray energy, $E^\mathrm{tot}_\gamma$, as a function of total fragment kinetic energy, $\rho (E_\gamma^\mathrm{tot},\mathrm{TKE})$, (a) before neutron emission and (b) after neutron emission for $^{252}$Cf(sf).  Black and red dashed lines in (a) outline the structures seen.  (c) Distribution of the total neutron energies, $E_n^\mathrm{tot}$, subtracted from the $Q$-values for $^{252}$Cf(sf).  Red diagonal and black horizontal hashed regions correspond to the events enclosed by the dashed lines in (a) with the same color.}
	\label{fig:TKEEgammatot}
\end{figure*} 

In Fig. \ref{fig:TKEEgammatot}, we show the calculated distribution $E^\mathrm{tot}_\gamma$ vs. (a) TKE$_\mathrm{pre}$ and (b) TKE$_\mathrm{post}$ for $^{252}$Cf(sf).  Distinct structures are clearly visible in both panels, which can be understood as correlations due to energy conservation.  These structures can be easily explained when only one fission event is considered where decreasing TKE$_\mathrm{pre}$ directly corresponds to an increase in TXE.  For the de-excitation of a fission fragment, a small increase in the initial excitation energy (decrease in the kinetic energy) leads to an increase in $E_\gamma^\mathrm{tot}$ since this change will not be enough to raise the excitation energy above the neutron separation energy, $S_n$, in the last daughter fragment.  However, when the excitation energy increases enough, this same daughter fragment will now be above $S_n$, another neutron can be emitted, and less excitation energy is now available for $\gamma$-ray decay, lowering $E_\gamma^\mathrm{tot}$.  

These structures also appear in the differences between $Q$ and $E_n^\mathrm{tot}$ for each fission event which is shown in Fig. \ref{fig:TKEEgammatot}(c).  Although there is a large spread of these values, specific peaks can still be seen, indicated by the red and black overlaid distributions which correspond to the events within the red and black dashed lines in panel (a).  The distributions of the $Q$-values and outgoing neutron energies are smooth, so these peaks arise from the neutron separation energies of the fission fragments.

\begin{figure*}
	\centering
	\includegraphics[width=\textwidth]{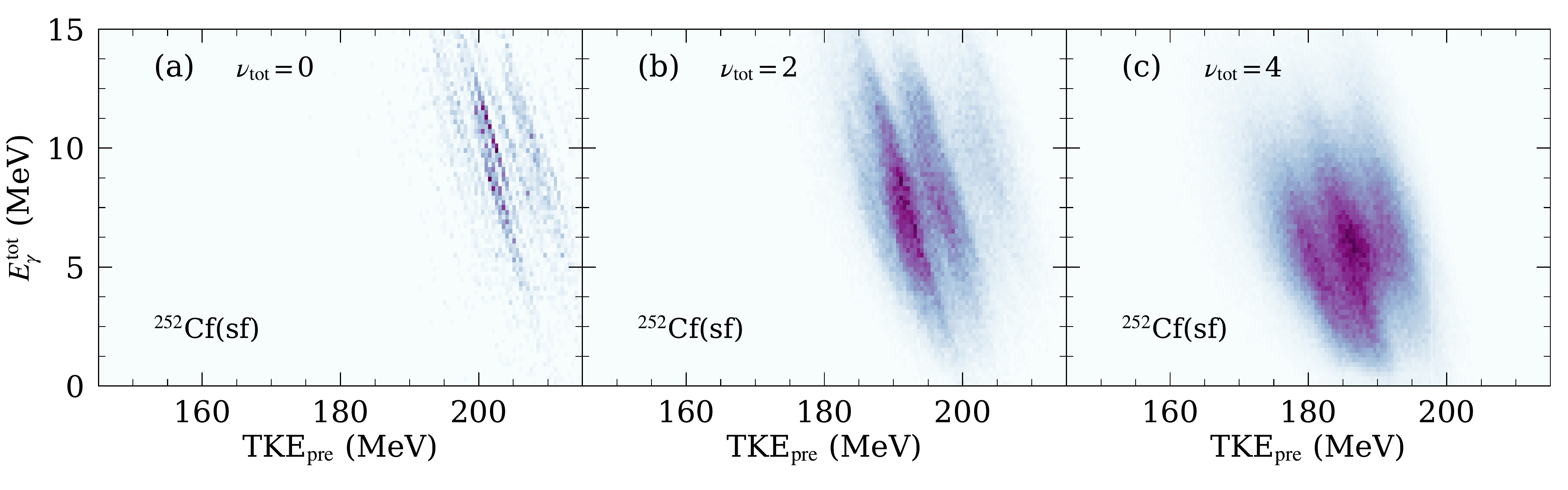}
	\caption{(Color online) Total $\gamma$-ray energy, $E_\gamma^\mathrm{tot}$, as a function of total fragment kinetic energy, TKE$_\mathrm{pre}$, $\rho (E_\gamma^\mathrm{tot},\mathrm{TKE} | \nu_i)$, for $^{252}$Cf(sf) for events where the total number of prompt neutrons emitted, $\nu_i$, is (a) zero (cold fission), (b) two, and (c) four.}
	\label{fig:nutotComponents}
\end{figure*}

These correlations between $E_\gamma^\mathrm{tot}$ and TKE also appear when we consider a fixed value of the total number of neutrons emitted during a fission event, $\nu_\mathrm{tot}$.  For events where no neutrons are emitted (cold fission), the energy conservation equation becomes $E^\mathrm{tot}_\gamma= Q - \mathrm{TKE}_\mathrm{pre}$.  All of the available excitation energy is dissipated through photon emission, and the features in the $E_\gamma^\mathrm{tot}$-TKE$_\mathrm{pre}$ plots correspond directly to the $Q$-values present in these reactions, Fig. \ref{fig:nutotComponents} (a).  For events where $\nu_\mathrm{tot}>0$, the negative correlations between $E_\gamma^\mathrm{tot}$ and TKE$_\mathrm{pre}$ still exist, but the structures shift toward lower TKE values.  In addition, because the variance of $E^\mathrm{tot}_n$ increases with increasing neutron multiplicity, the patterns are less clearly separated when $\nu_\mathrm{tot}$ increases.  This is illustrated in Fig. \ref{fig:nutotComponents}, panels (b) and (c), where we show $E_\gamma^\mathrm{tot}$ vs. TKE$_\mathrm{pre}$ for $\nu_\mathrm{tot}=2$ and $\nu_\mathrm{tot}=4$.  



The full $E_\gamma^\mathrm{tot}$-TKE distribution, $\rho (E_\gamma^\mathrm{tot},\mathrm{TKE})$, can be constructed as a superposition of the contribution from each $\nu_\mathrm{tot}$, $\rho (E_\gamma^\mathrm{tot},\mathrm{TKE} | \nu = \nu_\mathrm{tot})$, with weights given by $P(\nu=\nu_\mathrm{tot})$,
\begin{equation}
\rho (E_\gamma^\mathrm{tot},\mathrm{TKE}) = \sum \limits _{\nu_{i} =0} ^{\nu_\mathrm{max}} P(\nu_i) \rho (E_\gamma^\mathrm{tot},\mathrm{TKE} | \nu_i).
\end{equation}

\noindent Assuming the $\nu$-specific distributions $\rho (E_\gamma^\mathrm{tot},\mathrm{TKE}|\nu_i)$ can be reasonably calculated from a physics model where the fission fragment yields, energy sharing, and nuclear structure information are constrained by experimental measurements and that the total $\rho (E_\gamma^\mathrm{tot},\mathrm{TKE})$ distribution can be measured, then the coefficients $P(\nu_i)$ could be inferred with reasonable accuracy.

\section{Results and Discussion}

We first test this method using calculations from {\CGMF} to construct the $E_\gamma^\mathrm{tot}$-TKE correlation plots for each value of $\nu_\mathrm{tot}$ and then perform a gradient descent minimization of the $\chi^2$ to extract $P(\nu)$ from the full $E_\gamma^\mathrm{tot}$-TKE distribution.  The initial condition for $P(\nu)$ was a uniform distribution across $\nu$, but the results are insensitive to changes in this initial condition.  From the minimization, we are able to calculate uncertainties due to the fitting procedure from the square root of the diagonal elements of the covariance matrix of the extracted $P(\nu)$ values.  The covariances are calculated numerically around the best-fit value for $P(\nu)$.

To quantify the quality of the extracted neutron multiplicity distribution, we calculate the relative error on the first three factorial moments of $\nu$ ($\langle \nu \rangle$, $\langle \nu (\nu-1)\rangle $, $\langle \nu (\nu-1)(\nu-2)\rangle $), as
\begin{equation}
\varepsilon_1 = \frac{|\overline{\nu}^\mathrm{CGMF}-\overline{\nu}^\mathrm{fit}|}{\overline{\nu}^\mathrm{CGMF}},
\label{eq:chi2}
\end{equation}

\noindent where $\overline{\nu}^\mathrm{CGMF}$ is the average prompt neutron multiplicity from {\CGMF}, and $\overline{\nu}^\mathrm{fit}$ is that resulting from the fitted distribution.  $\varepsilon_2$ and $\varepsilon_3$ are defined likewise for $\langle \nu (\nu-1)\rangle $ and $\langle \nu (\nu-1)(\nu-2)\rangle $.  


To take into account the experimental resolution on TKE, we folded the TKE$_\mathrm{pre}$ values from {\CGMF} with a Gaussian of width $\delta_\mathrm{TKE}$.  Experimentally, the resolution is $\sim$1 MeV at best, and, in this work, we study values up to 4.5 MeV to mimic this effect and study its impact on our results.



Furthermore, due to experimental calibration, it is possible for the kinetic energies to be systematically shifted by up to a few MeV.  For this reason, we also tested the ability of our minimization routine to determine a systematic shift in TKE, $\Delta_\mathrm{TKE}$.  We defined a grid in TKE and performed the minimization procedure.  Whichever grid point resulted in the lowest residuals between the true $E_\gamma^\mathrm{tot}$-TKE distribution and the reconstructed distribution corresponded to the true value of $\Delta_\mathrm{TKE}$.

\begin{figure}
	\centering
	\includegraphics[width=0.45\textwidth]{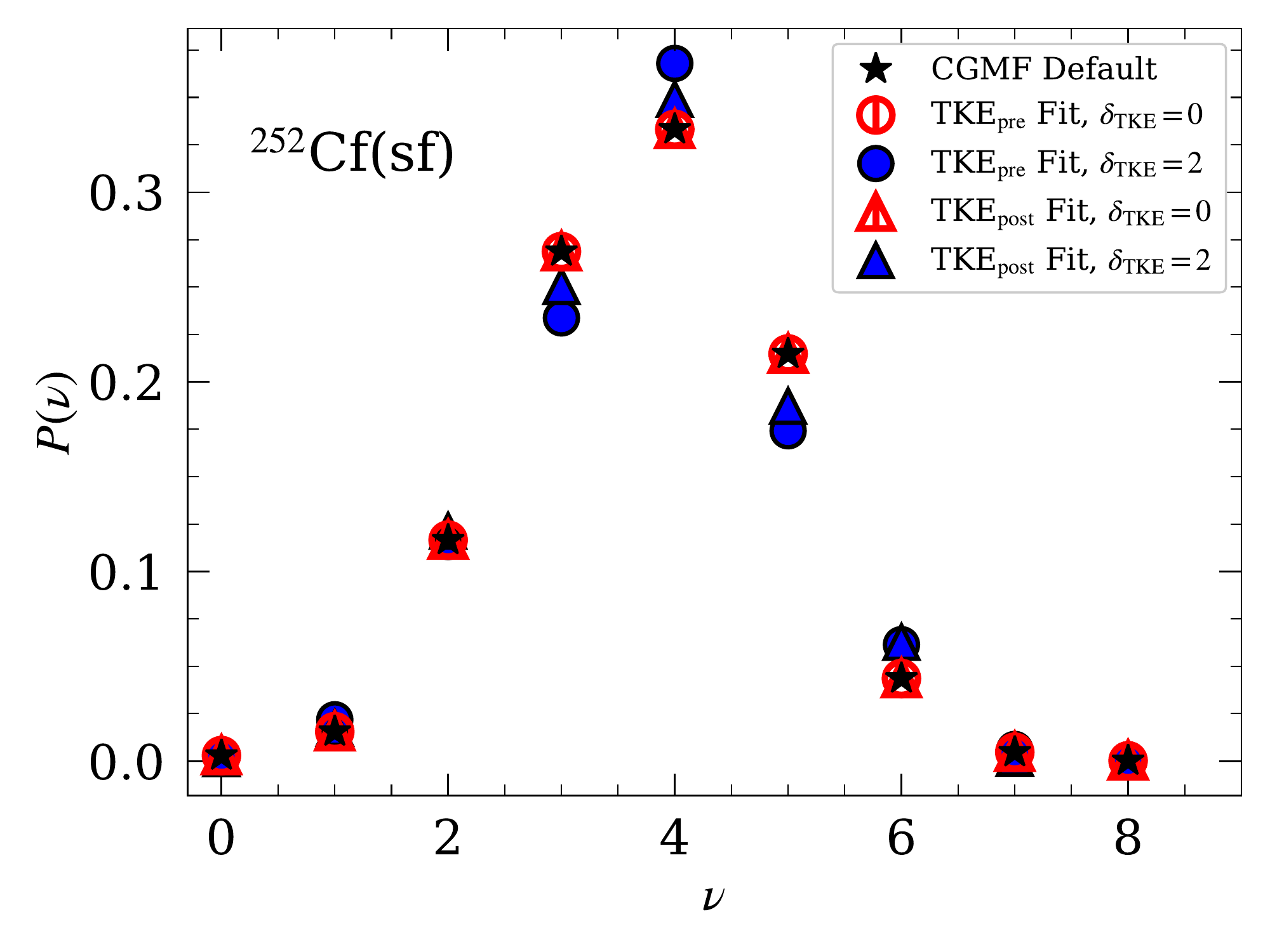}
	\caption{(Color online) For $^{252}$Cf (sf), $P(\nu)$ from {\CGMF} (black stars) compared to that extracted from the fitting procedure using TKE$_\mathrm{pre}$ (red open circles), TKE$_\mathrm{post}$ (red open triangles) both with $\delta_\mathrm{TKE}=0$ MeV, and for TKE$_\mathrm{pre}$ (blue filled cirles) and TKE$_\mathrm{post}$ (blue filled triangles) with $\delta_\mathrm{TKE}=2$ MeV.}
	\label{fig:252CfPnu}
\end{figure}

When no energy resolution or shift is considered, $\varepsilon_1=4\cdot 10^{-6}$, $\varepsilon_2=3\cdot 10^{-7}$, $\varepsilon_3=4\cdot 10^{-5}$, indicating that the extracted $P(\nu)$ is nearly identical to the reference value calculated in {\CGMF}.  This is shown in Fig. \ref{fig:252CfPnu} as the red open circles for the extracted $P(\nu)$, compared to the {\CGMF} reference value, black stars.  The error bars on the extracted values of $P(\nu)$ are the square roots of the diagonal values of the covariance matrix, as discussed in the beginning of this section, but are smaller than the size of the points in all cases.  As the energy resolution worsens, the quality of the extracted $P(\nu)$ declines, shown for $\delta_\mathrm{TKE}=2$ MeV in Fig. \ref{fig:252CfPnu}, blue filled circles.  Regardless of the shape of the extracted $P(\nu)$, the resulting $\overline{\nu}$ is determined within 1.5\% of the nominal value from {\CGMF}, independent of $\delta_\mathrm{TKE}$.  For the higher moments, the relative uncertainty increases more quickly with increasing $\delta_\mathrm{TKE}$, although the first three moments stay within 10\% of the nominal values from {\CGMF} until $\delta_{TKE}=4.0$ MeV.  Even for $\delta_\mathrm{TKE}=2$ MeV, the resulting $P(\nu)$ is similar enough to the true value (within $\sim 20\%$) that this type of extraction would be greatly beneficial in regions without any direct measurement of $P(\nu)$.

The emission of prompt neutrons smears the distribution in total kinetic energy and shifts it to lower fragment energies.  This is evident by comparing panel (b) to panel (a) in Fig. \ref{fig:TKEEgammatot}.  The distinct structures are almost completely removed in the full distribution, as seen here, as well as in the $\nu_\mathrm{tot}$ components, which are likewise shifted to lower TKE values.  The same analysis is repeated using TKE$_\mathrm{post}$.  Surprisingly, for $\delta_\mathrm{TKE}=0$ MeV, the extraction of $P(\nu)$ is similar to that using TKE$_\mathrm{pre}$, the red open triangles in Fig. \ref{fig:252CfPnu} compared to the red open circles.  The relative uncertainty for the first three factorial moments with TKE$_\mathrm{post}$ is lower than those calculated using TKE$_\mathrm{pre}$, especially as $\delta_\mathrm{TKE}$ increases.  For TKE$_\mathrm{post}$, the energy carried away by the neutrons leads to greater separation between the $\rho (E_\gamma^\mathrm{tot},\mathrm{TKE} | \nu_i)$ distributions.

Although we expect the total kinetic energy resolution to be the limiting factor in this type of measurement, we can also study how a resolution on the total $\gamma$-ray energy, $\delta_{\mathrm{E}_{\gamma}^\mathrm{tot}}$ affects the extracted $P(\nu)$.  We study $\delta_{\mathrm{E}_{\gamma}^\mathrm{tot}}$ values below 2 MeV.  Below $\delta_{\mathrm{E}_{\gamma}^\mathrm{tot}}=0.5$ MeV, there is no difference in the extracted $P(\nu)$ distribution compared to the distribution calculated directly from {\CGMF}.  Even at $\delta_{\mathrm{E}_{\gamma}^\mathrm{tot}}=1.0$ MeV, the quality of the extracted $P(\nu)$ is similar to the distribution extracted when $\delta_\mathrm{TKE}=2.0$ MeV.

Because the TKE distribution is particularly important for reproducing $P(\nu)$, it is worth investigating whether $P(\nu)$ could be extracted in a similar fashion just using the TKE distributions, $Y(\mathrm{TKE})=\sum \limits_{\nu_{i} =0} ^{\nu_\mathrm{max}} P(\nu_i) Y(\mathrm{TKE}|\nu_i)$.  Indeed, without considering any resolution on TKE, $P(\nu)$ can be extracted exactly from $Y(\mathrm{TKE})$.  However, as the TKE resolution worsens, the quality of the extracted $P(\nu)$ degrades quickly; for $\delta_\mathrm{TKE}=2.0$ MeV, the distribution is no longer peaked at $\nu_\mathrm{tot}=4$.  In this case, $E_\gamma^\mathrm{tot}$ provides an additional constraint in the fitting procedure leading to a more reliable shape of $P(\nu)$ even as the energy resolution worsens.

When extracting $P(\nu)$ from experimental data, the quality of the extraction depends on how robust the multiplicity and $E_\gamma^\mathrm{tot}$-TKE distributions are to changes in the underlying fission model.  Ultimately, model uncertainties should be treated in a rigorous manner (e.g. Bayesian methods, model-form uncertainty), but while this framework is being developed, we investigate these uncertainties by varying parts of the model within {\CGMF}, constructing the $\rho (E_\gamma^\mathrm{tot},TKE | \nu_i)$ components from the modified versions of {\CGMF}, and then extracting $P(\nu)$ from $\rho (E_\gamma^\mathrm{tot},TKE)$ calculated with the nominal version of the code.  We change the model for the yields in mass and TKE, the spin distribution parameter, and the excitation energy sharing between the daughter fragments.  The initial Gaussian parameterization for $Y(A)$ and $Y(TKE|A)$ in {\CGMF} are replaced with an optimized version of the Brosa mode parameterization \cite{Brosa1990} which does not use a Gaussian for $Y(TKE|A)$ as in {\CGMF}.  The spin parameter, $\alpha$, was changed from the default value of 1.7 to 1.6 and 1.5, and $R_T$ was changed from a function of mass to fixed values of 1.1 and 1.2.

\begin{figure}
\centering
\includegraphics[width=0.45\textwidth]{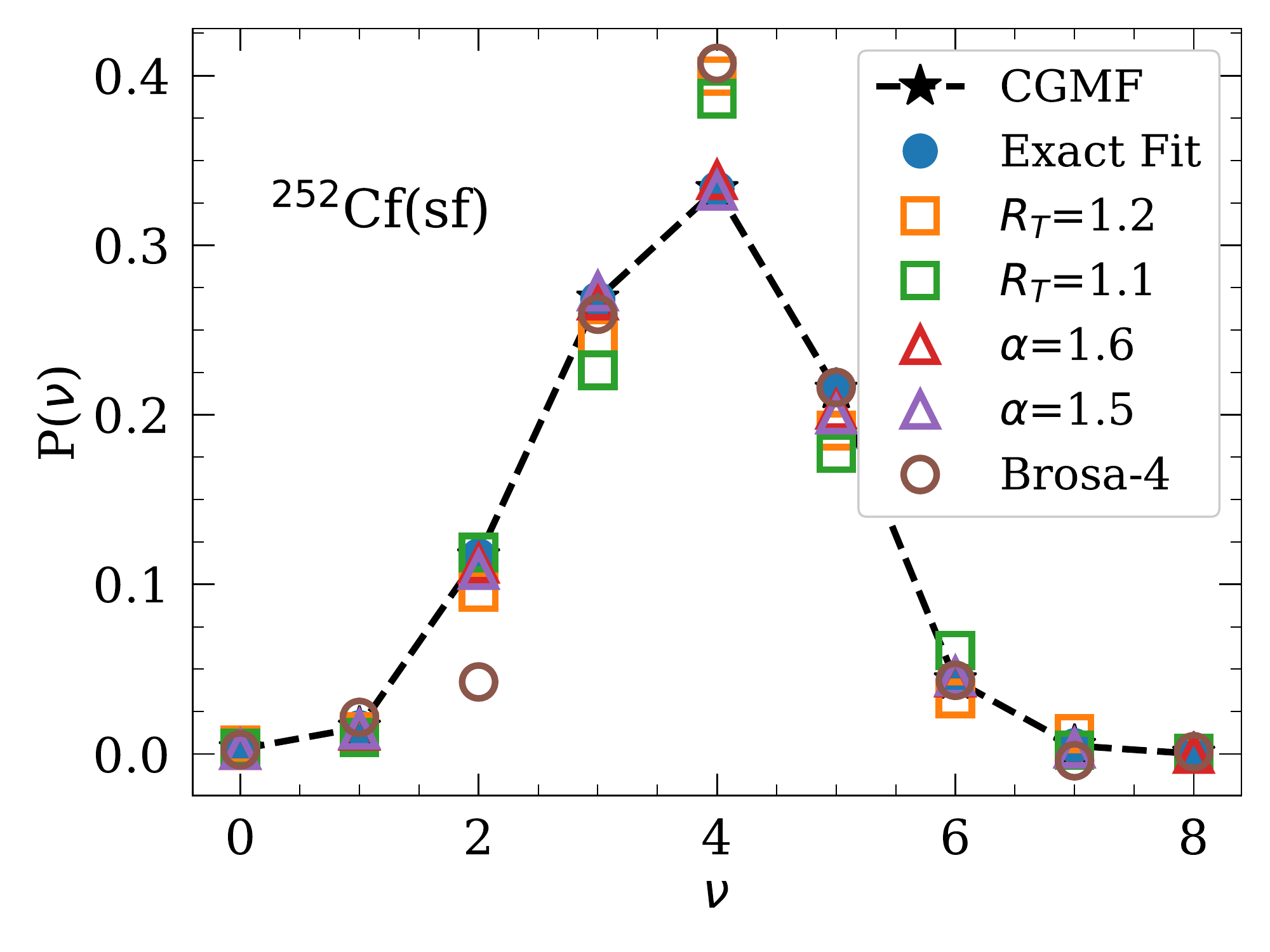}
\caption{(Color online) Comparison of $P(\nu)$ from {\CGMF} for $^{252}$Cf(sf), black stars, to the extracted $P(\nu)$ when changes are made to the models within {\CGMF}.  Open orange (green) squares correspond to $R_T=1.2$ (1.1), open red (purple) triangles to $\alpha=1.6$ (1.5), and open brown circles to the Brosa parameterization of $Y(A,TKE)$.  Full blue circles show $P(\nu)$ extracted from the default version of {\CGMF}.}
\label{fig:models}
\end{figure}

The extracted $P(\nu)$ from each of these model modifications is shown in Fig. \ref{fig:models}.  In each case, $P(\nu)$ varies by less than 25\% at the peak of the distribution while $\overline{\nu}$ ($\varepsilon_1$) varies by no more than 3\%.  The relative uncertainties $\varepsilon_2$ and $\varepsilon_3$ from these model changes are comparable to those introduced by $\delta_\mathrm{TKE}=2.0$ MeV.  The largest deviations from the original $P(\nu)$ comes when the Brosa model of $Y(A,TKE)$ is implemented, consistent with previous findings that the TKE distribution has a large impact on $P(\nu)$.

We note here that there are other contributions that could lead to model inaccuracies in {\CGMF}, including neutron and $\gamma$-ray input.  The neutron emission depends on the optical potentials that are used to calculate the transmission coefficients, which most likely have large uncertainties away from stability.  The $\gamma$-ray emission depends on the level densities and the strength functions.  The strength functions, particularly M1, have more uncertainty as there is limited experimental data to constrain these functions.  To fully understand the affect that these models have on the resulting {\CGMF} calculations would require a dedicated study, which, although interesting, is beyond the scope of this work.

The same studies are performed for neutron-induced fission reactions on $^{235}$U and $^{239}$Pu, both at thermal and 4.0 MeV incident neutron energy.  For thermal neutrons, the structures in the distributions of $E_\gamma^\mathrm{tot}$-TKE$_\mathrm{pre}$ are similar to that of $^{252}$Cf(sf).  As the energy of the incident neutron increases, these structures disappear, similar to the nearly featureless TKE$_\mathrm{post}$ distributions.  Here, the distributions in TKE are also well separated for the different values of $\nu_\mathrm{tot}$, and the same fitting procedure works for these reactions as well.  Except for $^{235}$U$(n,f)$, the first three moments are all within 10\% of the nominal values from {\CGMF} with $\delta_\mathrm{TKE}=2.0$ MeV.  For the two targets where we study neutron-induced fission, the relative differences for these moments are smaller for incident neutron energy of 4.0 MeV than for the thermal neutrons. 

\begin{figure}
	\centering
	\includegraphics[width=0.4\textwidth]{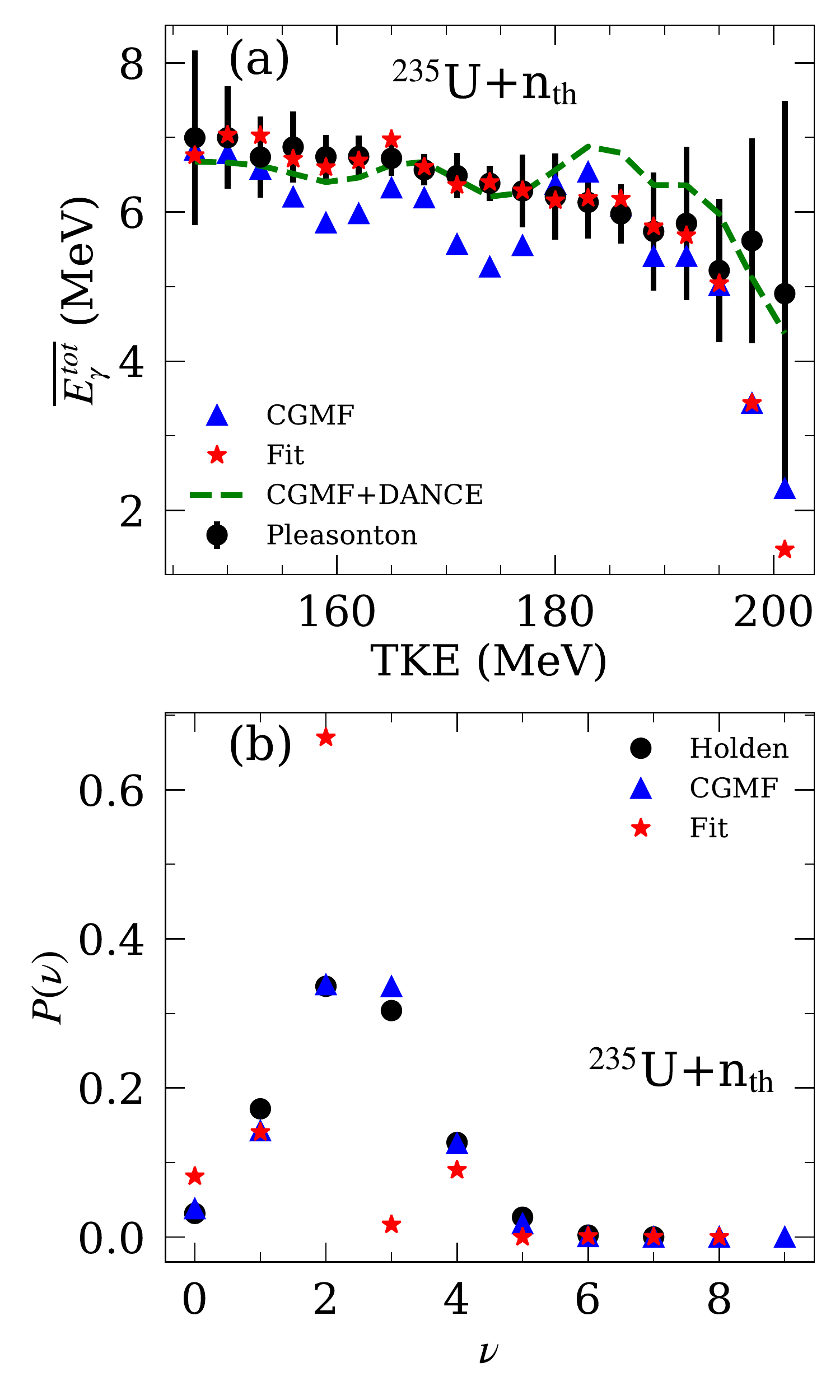}
	\caption{(Color online) (a) Comparison of $\overline{E_\gamma^\mathrm{tot}}$ as a function of TKE for $^{235}$U$+n_\mathrm{th}$ between the data of \cite{Pleasonton1972} (black circles), the default {\CGMF} calculations (blue triangles) and the the curve resulting from the $\chi^2$ minimization (red stars).  For reference, the {\CGMF} calculation folded with the DANCE detector response is shown, shifted up to match the magnitude of the data (green dashed line).  (b) Comparison of $P(\nu)$ for the same reaction between the default {\CGMF} calculation (blue triangles) and that extracted from the minimization routine (red stars).  Evaluated data from \cite{Holden1988} is shown for reference (black circles).}
	\label{fig:Pleasonton}
\end{figure}

Data measured at Los Alamos with DANCE (Detector for Advanced Neutron Capture Experiments) do exist for the full $E_\gamma^\mathrm{tot}$-TKE$_\mathrm{pre}$ distribution for $^{252}$Cf spontaneous fission \cite{Rusev2017}.  DANCE, a $4\pi$ $\gamma$-ray detector, was coupled with four silicon detectors in order to measure $\gamma$-ray energies in coincidence with the fission fragment kinetic energies.  These data show hints of the correlations seen in {\CGMF}, but the analysis has been hindered by poor resolution in TKE and $E_\gamma^\mathrm{tot}$.  However, for thermal neutrons incident on $^{235}$U, $\overline{E_\gamma^\mathrm{tot}}$ as a function of TKE was indirectly measured by Pleasonton, et al. \cite{Pleasonton1972}.  A similar fitting procedure to the one described here can be used to extract $P(\nu)$ from this data:  after the $E_\gamma^\mathrm{tot}$-TKE distribution is constructed, the average $\gamma$-ray energies are calculated for each TKE bin, then fit to the data using a $\chi^2$ minimization.  

For this observable, {\CGMF} calculations follow the same trend as the experimental data, although more structures are seen in the calculation, especially for TKE values within the range of 155 to 180 MeV, Fig. \ref{fig:Pleasonton} (a) blue triangles.  The fitting procedure flattens out these fluctuations (red stars in (a)).  However, these fluctuations appear to be related to $P(\nu=2,3)$, at the peak of the distribution, which now no longer agrees with the default {\CGMF} calculations or other experimental measurements, Fig. \ref{fig:Pleasonton} (b).  This discrepancy is most likely due to the direct comparison of theory with experimental data - without taking detector response into account.  Forward propagation of {\CGMF} calculations through the experimental response of DANCE (green dashed line, Fig. \ref{fig:Pleasonton} (a)) shows that structures present in {\CGMF} calculations for $\overline{E_\gamma^\mathrm{tot}}$(TKE) are mostly washed out when the detector response is included. Note that the folded {\CGMF} results have been shifted up to agree with the Pleasonton data at $\mathrm{TKE}\sim 170$ MeV; the cause of this shift is the simulated DANCE detector response that includes the experimental thresholds and energy resolution of the individual detectors and all material between the target and detectors, which decreases $E_\gamma^\mathrm{tot}$ by a few MeV.


\section{Conclusion}

In summary, we propose a novel method to extract the neutron multiplicity distribution from correlation plots of the total $\gamma$-ray energy and the total fission fragment kinetic energy, without measuring neutrons.  When no resolution in total $\gamma$-ray or kinetic energy is considered, $P(\nu)$ can be extracted within 25\% of the nominal value considering model uncertainties.  When a resolution in total kinetic energy is applied, $P(\nu)$ can still be extracted reliably with a resolution of 2.0 MeV, at which point the uncertainty from the resolution outweighs the model uncertainty.  We also applied this method to a measurement for thermal neutron-induced fission for $^{235}$U with $\overline{E_\gamma^\mathrm{tot}}$ as a function of TKE.  Although there are discrepancies between the extracted $P(\nu)$ and available experimental data, our analysis suggests that this discrepancy is likely due to detector resolution effects that could not be taken into account during the fitting.  New measurements of the correlations between the total $\gamma$-ray and fragment kinetic energies are encouraged to better validate this novel method.



\begin{acknowledgements}
	This work was performed under the auspice of the U.S. Department of Energy by Los Alamos National Laboratory under Contract 89233218CNA000001 and was supported by the Office of Defense Nuclear Nonproliferation Research \& Development (DNN R\&D), National Nuclear Security Administration, U.S. Department of Energy.  We gratefully acknowledge the support of the U.S. Department of Energy through the LANL/LDRD Program and the Center for Non Linear Studies.
\end{acknowledgements}

\bibliography{EgtotTKE}

\end{document}